\documentclass[conference]{IEEEtran}
\IEEEoverridecommandlockouts
\usepackage{cite}

\usepackage{amsmath,amssymb,amsfonts}
\usepackage{algorithmic}
\usepackage{graphicx}
\usepackage{csquotes}
\usepackage{textcomp}
\usepackage[table]{xcolor}

\usepackage{colortbl}
\usepackage{array,tabularx}
\definecolor{tablerow1}{RGB}{230,230,230}
\definecolor{tablerow2}{RGB}{245,245,245}

\usepackage[most]{tcolorbox}

\def\BibTeX{{\rm B\kern-.05em{\sc i\kern-.025em b}\kern-.08em
    T\kern-.1667em\lower.7ex\hbox{E}\kern-.125emX}}
\begin{document}

\title{\huge{ChatGPT Prompt Patterns for Improving Code Quality, Refactoring, Requirements Elicitation, and Software Design

}}

\author{
\IEEEauthorblockN{Jules White, Sam Hays, Quchen Fu, Jesse Spencer-Smith, Douglas C. Schmidt}
\IEEEauthorblockA{\textit{Department of Computer Science, Vanderbilt University, Nashville, TN, USA} \\
\{jules.white, quchen.fu, george.hays, jesse.spencer-smith, douglas.c.schmidt\}@vanderbilt.edu}
}

\maketitle

\begin{abstract}
This paper presents prompt design techniques for software engineering, in the form of patterns, to solve common problems when using large language models (LLMs), such as ChatGPT to 
automate common software engineering activities,
such as ensuring code is decoupled from third-party libraries and creating an API specification from a requirements list. This paper provides two contributions to research on using LLMs for software engineering.  First, it provides a catalog of patterns for software engineering that classifies patterns according to the types of problems they solve. Second, it explores several prompt patterns that have been applied to improve requirements elicitation, rapid prototyping, code quality, deployment, and testing. 
\end{abstract}

\begin{IEEEkeywords}
large language models, prompt patterns, prompt engineering, software engineering
\end{IEEEkeywords}

\section{Introduction}
\label{sec:intro}

\textbf{Overview of LLMs and prompts for automating software engineering tasks}.
Large-scale language models (LLMs)~\cite{bommasani2021opportunities} are rapidly being adopted by software developers and applied to generate code and other artifacts associated with software engineering. Popular examples of LLM-based tools applied for these purposes include  ChatGPT~\cite{bang2023multitask} and GitHub Copilot~\cite{github}. Initial research indicates that such chat-enabled artificial intelligence (AI) tools can aid a range of common software development and engineering tasks~\cite{carleton2022architecting}. 

A key to the adoption of these tools has been the creation of LLMs and IDE-integrated services around them. Any user can access these complex LLM capabilities by simply typing a message to ChatGPT or opening popular integrated development environments (IDEs)~\cite{github,asare2022github,pearce2022asleep}, such as IntelliJ~\cite{krochmalski2014intellij} and Visual Studio Code. Previously, leveraging these capabilities required substantially more time and effort. In addition, prior state-of-the-art LLMs were not widely accessible to users.  

Interacting with an LLM in general involves feeding it  "prompts"~\cite{liu2023pre}, which are natural language instructions used to provide context to the LLM and guide its generation of textual responses. In a chat-based environment, a prompt is a chat message that the user sends to an LLM, such as ChatGPT. The remainder of this paper focuses on the ChatGPT chat-enabled LLM. 

In the context of software engineering, a prompt is a natural language instruction given to an LLM to facilitate its generation of requirements, code and software-related artifacts (such as documentation and build scripts), as well as to simulate certain aspects of a software system. Prompts are thus a form of programming used to instruct an LLM to perform software engineering tasks.  For example, in an IDE environment (such as 
Copilot~\cite{github}), a prompt can be a comment, method definition, or source file. 

One way to use prompts is to directly ask an LLM to provide some information or generate some code. Another use of prompts is to dictate rules for the LLM to follow going forward, such as conforming to certain coding standards. Both types of prompts program the LLM to accomplish a task. The second type of prompt, however, customizes future interactions with the LLM by defining additional rules to follow or information to use when performing subsequent tasks. We cover both types of patterns in this paper.

\textbf{Overview of prompt patterns for software engineering tasks}.
 This paper builds on our prior work that introduced the concept of \textit{prompt patterns}~\cite{white2023prompt}, which are reusable prompt designs to solve problems in LLM interaction. Similar to software patterns~\cite{gamma1995design,schmidt2013pattern}, prompt patterns codify sound design experience, thereby providing a reusable solution to common problems in LLM interaction, such as ensuring that programs adhere to certain design principles or secure coding guidelines.
 
 Developers and software engineers can use prompt patterns to establish rules and constraints that improve software quality attributes (such as modularity or reusability) when working with LLMs. For example, prompt patterns can ensure that generated code (or user-provided code being refactored) separates business logic from code with side-effects (e.g., file system access, database access, network communication, etc.). These types of constraints make business logic easier to test and reason about since it is decoupled from harder-to-test and harder-to-understand side-effecting code. Prompt patterns can also require that third-party libraries have intermediate abstractions inserted between the libraries and the business logic depending on them to ensure the code is not tightly-coupled to external dependencies.

\textbf{Towards a prompt pattern catalog for software engineering.} This paper extends our prior 
work~\cite{white2023prompt} by focusing on creating a catalog of prompt patterns that can be applied collaboratively throughout the software life-cycle. We introduce a variety of prompt patterns in this paper, ranging from patterns that simulate and reason about systems early in the design phase to patterns that help alleviate issues with LLM token limits when generating code. In addition, we explore relationships between patterns by examining patterns compounds and sequences that are most effective when employed in combination with each other.

The remainder of this paper is organized as follows: Section~\ref{sandf} gives an overview of prompt pattern structure and functionality; Section~\ref{impl} introduces the catalog of prompt patterns covered in the paper; Section~\ref{arch} describes prompt patterns used during requirements elicitation and system design; Section~\ref{qual} describes prompt patterns that help LLMs generate higher quality code and refactor human-produced code; Section~\ref{related} compares our research on prompt patterns with related work; and Section~\ref{conclusion} presents concluding remarks and lessons learned.

 \section{Prompt Pattern Structure and Functionality}
\label{sandf}
Prompt patterns are documented using a similar structure to software patterns, with analogous versions of the name, classification, intent, motivation, structure, example implementation, and consequences. Each of these sections for the prompt pattern form is described briefly below:\footnote{Our prior work~\cite{white2023prompt} defines the fundamental structure of a prompt pattern and compares it with software patterns. We briefly define prompt patterns for completeness below, but we refer the reader to our prior work for additional details.}

\begin{itemize}
    \item \textbf{A name and classification}. 
    The name provides a unique identifier for the pattern that can be referenced in discussions and the classification groups the pattern with other patterns based on the types of problems they solve. The classification used in this paper is shown in Table~\ref{class:table}.
    \item \textbf{The intent and context} captures the problem that the pattern solves and the goals of the pattern.
    \item  \textbf{The motivation} explains the rationale and importance of the problem that the pattern is solving.  
    \item  \textbf{The structure and key ideas}. The structure describes the fundamental contextual information that needs to be provided by the LLM to achieve the expected behavior. These ideas are listed as a series of statements, but can be reworded and adapted by the user, as long as the final wordings convey the key information. 
    \item  \textbf{Example implementation} shows specific implementations of the pattern and discusses them.
    \item \textbf{Consequences} discusses the pros and cons of using the pattern and discussion of how to adapt the pattern for different situations.
\end{itemize}

Prompt patterns can take various forms. In the context of patterns that enable LLMs to perform software engineering tasks, a prompt typically starts with a conversation scoping statement, such as "from now on", "act as a X", "for the next four prompts", etc. These statements dictate to the LLM that it must change its operation going forward based on the prompt being provided. For example, the following prompt pattern is an adaptation of the \textit{Output Automater} pattern~\cite{white2023prompt} that uses "from now on" to automate production of a list of dependencies for generated code:

\begin{displayquote}
 "from now on, automatically generate a python requirements.txt file that includes any modules that the code you generate includes." 
 \end{displayquote}

After the initial conversational scoping statement, the prompt includes a number of statements that provide the ground rules  the LLM should follow in output generation and prompt flow for software engineering tasks. These output rules may include one or more conditional statements indicating when specific rules should be applied. For example, the following prompt:

\begin{displayquote}
    "From now on, whenever you generate code that spans more than one file, generate a python script that can be run to automatically create the specified files or make changes to existing files to insert the generated code."
\end{displayquote}

Normally, a user must manually open and edit multiple files to add generated code that spans multiple files to a project. With the above prompt, ChatGPT will generate a script to automate opening and editing each file for the user and eliminate potential manual errors. The prompt is scoped to "from now on" and then includes a conditional "whenever you generate code that spans more than one file", followed by the rule to generate a python script. This prompt form is an example of the \textit{Output Automator} pattern from ~\cite{white2023prompt}, applied to software engineering.

\section{A Catalog of Prompt Patterns for Automating Software Engineering Tasks}
\label{impl}

This section summarizes our catalog of 13 prompt patterns that have been applied to solve common problems in the domain of conversational LLM interaction and output generation for automating common software tasks. We partitioned these 13 prompt patterns into four categories to help users navigate and apply these patterns more effectively. Table~\ref{class:table} outlines the initial classifications for the catalog of prompt patterns for automating software engineering tasks identified by our work with ChatGPT.

\begin{table}[htbp]
\caption{Classifying Prompt Patterns for Automating Software Engineering Tasks}
\centering
\begin{tabular}{|l|l|}
\hline
\textbf {Requirements Elicitation} & Requirements Simulator \\
 & Specification Disambiguation \\ 
 & Change Request Simulation \\
\hline
\textbf {System Design and Simulation} & API Generator \\
 & API Simulator \\
 & Few-shot Example Generator \\
 & Domain-Specific Language (DSL) Creation \\
 & Architectural Possibilities \\
\hline
\textbf {Code Quality} & Code Clustering \\
 & Intermediate Abstraction \\
 & Principled Code \\
 & Hidden Assumptions \\ 
\hline
\textbf {Refactoring} & Pseudo-code Refactoring \\
& Data-guided Refactoring \\
\hline
\end{tabular}
\label{class:table}
\end{table}

Two areas of LLM usage in software engineering that have received scant attention thus far include (1) requirements elicitation and (2) system design and specification. These areas represent some of the most important aspects of software engineering, however, and commonly yield changes late in the development cycle that cause schedule overruns, unanticipated costs, and risk. The \textit{Requirements Elicitation} patterns listed in Table~\ref{class:table} aid in creating requirements and exploring their completeness with respect to desired system capabilities and accuracy. Other patterns in this category use the LLM serve as a trusted intermediary to reason about the impact of changes. 

The \textit{System Design \& Simulation Patterns} patterns category listed in Table~\ref{class:table} explores patterns that address issues creating concrete design specifications, domain-specific languages, and exploring alternative architectures. The section demonstrates ways to simulate aspects of a system to help identify deficiencies early in the life-cycle, \textit{i.e.}, when they are less costly and disruptive to remediate. 

There is considerable concern over the quality of code produced by LLMs, as well as written by humans~\cite{pearce2022asleep, asare2022github, borji2023categorical}. The \textit{Code Quality} patterns category introduces several patterns that improve both LLM and human-generated code. LLMs can often reason effectively about abstraction, as well as generate relatively modular code. The patterns listed in this category in Table~\ref{class:table} help ensure certain abstraction and modularity attributes are present in code, \textit{e.g.}, they facilitate replacement of third-party libraries by introducing an interface between them and business logic. 

Finally, the \textit{Refactoring} patterns listed in Table~\ref{class:table} provide various means to effectively refactor code using LLMs. LLMs like ChatGPT have a surprisingly powerful understanding of abstract coding constructs, such as pseudo-code. Innovative approaches to refactoring are therefore discussed to allow specification of refactoring at a high-level, such as using pseudo-code to describe code structure.

All examples in this paper were tested with the ChatGPT LLM. Our process for identifying and documenting these patterns combined exploring community-posted prompts on the Internet and creating independent prompts to support our own software engineering work with ChatGPT. Our objective is to codify a catalog of software engineering prompt patterns that can be easily adapted or reused for other LLMs, much like how classic software patterns can be implemented independently in various programming languages. 


\section{System Requirements \& Architecture Patterns}
\label{arch}

This section describes prompt patterns used during requirements elicitation and system design.


\subsection{The Requirements Simulator Pattern}

\subsubsection{Intent and Context}

This pattern allow stakeholders to explore the requirements of a software-reliant system interactively to determine if certain functionality is captured properly. The simulation output should provide additional details regarding the initial requirements and new requirements added to accomplish the tasks the stakeholders tried to perform in the simulation. The goal of this pattern is to aid in elicitation and analysis of the completeness of software requirements. 

\subsubsection{Motivation} 

Changes late in a software system's development are generally more expensive to remediate than early in the development phase. Unfortunately, many requirement changes are made late in the development cycle when they are more costly to fix. A common source of issues with requirements is that the requirements do not adequately describe the needs of the system. The motivation of this pattern is to use an LLM to simulate interactions with the system based on descriptions of the tasks that a user might want to perform and identify missing requirements.

\subsubsection{Structure and Key Ideas}

The fundamental contextual statements are as follows:

\rowcolors{2}{tablerow1}{tablerow2}
\begin{tcolorbox}[enhanced, notitle, clip upper,
       tabularx*={\renewcommand*{\arraystretch}{1.4}}%
                 {>{\raggedright\arraybackslash\hsize=0.1\hsize}X%
              >{\raggedright\arraybackslash\hsize=1.9\hsize}X}]
    \arrayrulecolor{white}
    \textbf{~} & \textbf{Requirements Simulator Pattern} \\ \hline
      1. & I want you to act as the system \\ \hline
      2. & Use the requirements to guide your behavior \\ \hline
      3. & I will ask you to do X, and you will tell me if X is possible given the requirements. \\ \hline
      4. & If X is possible, explain why using the requirements. \\ \hline
      5. & If I can't do X based on the requirements, write the missing requirements needed in format Y.    
\end{tcolorbox}

\subsubsection{Example Implementation}

A sample implementation of this pattern is shown below. The implementation focused on task-based exploration of the system's capabilities. The implementation specifically refines the format of the requirements to be user stories, so the LLM will produce requirements in the desired format. 

The prompt implementation assumes that the requirements have been given to the LLM prior to use of the prompt. The requirements could be typed in manually or generated by ChatGPT through a series of prompts asking for requirements related to a particular system. Any approach will work as long as the requirements are in the current context of the prompt. 

\begin{displayquote}
    "Now, I want you to act as this system. Use the requirements to guide your behavior. I am going to say, I want to do X, and you will tell me if X is possible given the requirements. If X is possible, provide a step-by-step set of instructions on how I would accomplish it and provide additional details that would help implement the requirement. If I can't do X based on the requirements, write the missing requirements to make it possible as user stories."
\end{displayquote}

An extension to this implementation is to include a screen-oriented exploration of the system. Whereas the prior example  focuses more on interrogating the system to see if a task is possible, the example below walks the user through individual screens. This approach of screen-by-screen walkthrough is similar to classic text-based adventure games, such as Zork. 

\begin{displayquote}
    "Now, I want you to act as this system in a text-based simulator of the system. Use the requirements to guide your behavior. You will describe the user interface for the system, based on the requirements, and what I can do on each screen. I am going to say, I want to do X, and you will tell me if X is possible given the requirements and the current screen. If X is possible, provide a step-by-step set of instructions how I would accomplish it and provide additional details that would help implement the requirement. If I can't do X based on the requirements, write the missing requirements to make it possible as user stories. Whenever the state of the user interface changes, update the user on what they are looking at.

    Tell me what I am looking at in the system and ask me what I want to do."
\end{displayquote}


\subsubsection{Consequences}

One consequence of this pattern is that a system can be simulated in greater levels of fidelity as the requirements are expanded. Since requirements can be rapidly expanded with this pattern through interactive exploration, the simulation of the system gains fidelity throughout the simulation. This iterative process allows rapid generation of large requirement sets.

A limitation of using the pattern in isolation is that the descriptions of the system are purely textual, which may provide insufficient information to capture all requirements accurately. One way to overcome this limitation and help write better requirements and later generate design mockups is to combine this pattern with the \textit{Visualization Generator} pattern~\cite{white2023prompt} This combination of patterns enables simulation of the system, as well as possible visualizations of each screen generated by another AI tool, such as Dall-E, Midjourney, or Stable Diffusion. A modified   prompt that combines the \textit{Requirements Simulator} and \textit{Visualization Generation} patterns would be: 

\begin{displayquote}
    "...In addition to the textual screen description, provide a Dall-E prompt that I can use to generate wireframes of what the screen might look like.
\end{displayquote}

A sample output of combining these two patterns when used in an exploration of a web application with Twitter-like functionality for socially sharing ChatGPT prompts is shown below:

\begin{displayquote}
    \textbf{ChatGPT:} Sure, I can provide a DALL-E prompt to generate a wireframe for the screen description. Here is the DALL-E prompt for the screen where users can delete their prompts: A wireframe of the 'My Prompts' screen where users can see a list of their own prompts. Each prompt is displayed with its title and description. Next to each prompt is a 'Delete' button that users can click to remove the prompt from the social network. At the bottom of the screen, there is a 'Back' button that users can click to return to the previous screen.
\end{displayquote}

\begin{figure}
    \centering
    \includegraphics[width=9cm]{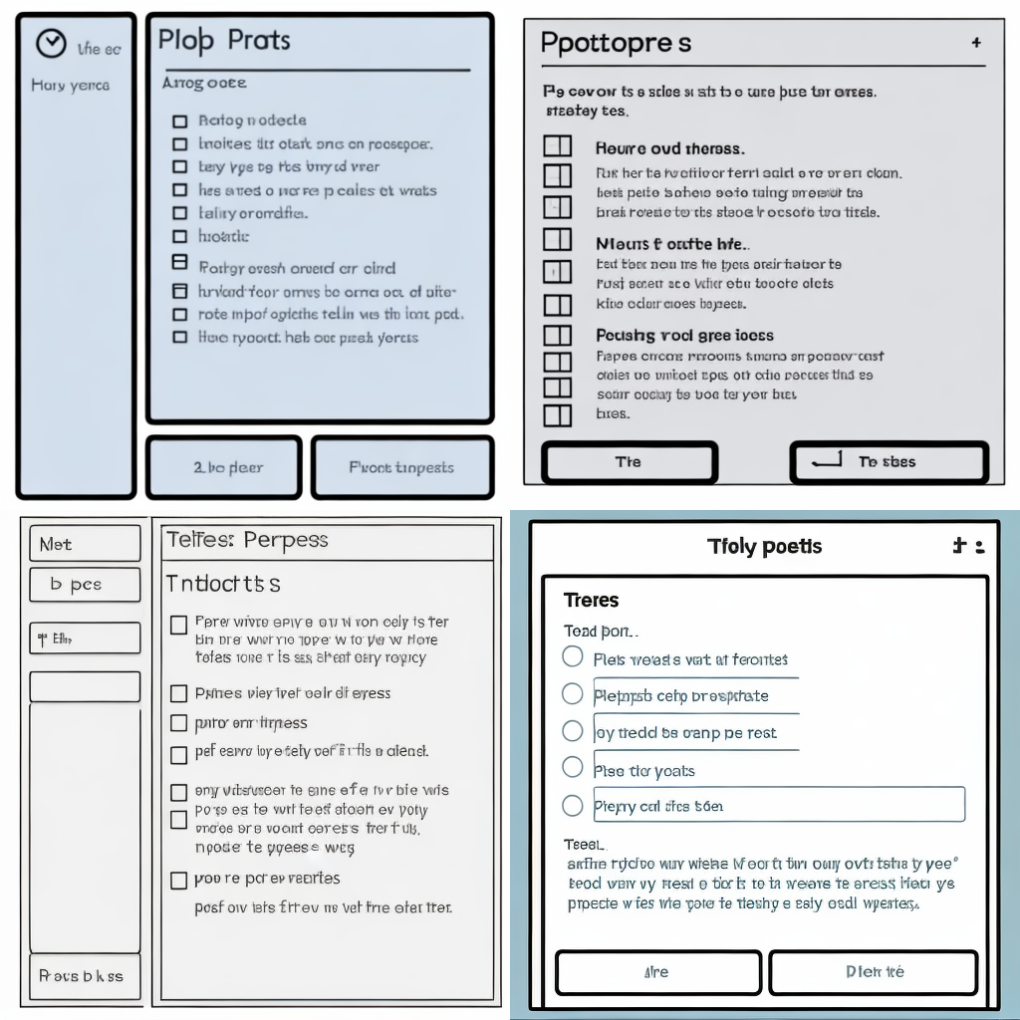}
    \caption{Wireframe Samples Generated from Combining the Requirements Simulator and Visualization Generator Patterns}
    \label{fig:mockup}
\end{figure}

The resulting wireframe generated with Midjourney rather than Dall-E is shown in Figure~\ref{fig:mockup}. This figure shows the generated image is a rough interpretation of possible realizations of the prompt. However, such visualizations  may be easier to reason about than textual descriptions, which can spur productive discussions between participants. In addition, the randomness in image generation may yield unexpected layouts or user interface approaches that may inspire new ideas.

\subsection{The Specification Disambiguation Pattern}

    \subsubsection{Intent and Context}
        This prompt pattern causes the LLM to review specifications provided to a developer or development team by non-technical or semi-technical personnel. The goal is to ensure any areas of potential miscommunication or ambiguity is caught early and clarified. 
  
    \subsubsection{Motivation}
        It is common for software developers to work with one or more product or project managers who generate requirements and provide those to the developers. Often these requirements stem from discussions with customers and there may be a loss of meaningful context in the capture and delivery of those requirements. This prompt pattern helps automate some of the disambiguation work that might be needed with the requirements, ideally before they are delivered to developers.
        
    \subsubsection{Structure and Key Ideas}

Fundamental contextual statements:

\rowcolors{2}{tablerow1}{tablerow2}
\begin{tcolorbox}[enhanced, notitle, clip upper,
       tabularx*={\renewcommand*{\arraystretch}{1.4}}%
                 {>{\raggedright\arraybackslash\hsize=0.1\hsize}X%
              >{\raggedright\arraybackslash\hsize=1.9\hsize}X}]
    \arrayrulecolor{white}
    \textbf{~} & \textbf{Specification Disambiguation Pattern} \\ \hline
      1. & Within this scope \\ \hline
      2. & Consider these requirements or specifications \\ \hline
      3. & Point out any areas of ambiguity or potentially unintended outcomes
\end{tcolorbox}

\subsubsection{Example Implementation}

Here we consider a scenario in which a non-technical product manager has consulted with customers and tried to capture the essence of their requirements. These requirements have been communicated in natural language unsuitable for software requirements due to inconsistent context (\textit{i.e.}, the \textit{what}, not the \textit{how}). The product manager may request additional areas of clarification by posing the following prompt to the LLM:
\begin{displayquote}
    ``The following will represent system requirements. Point out any areas that could be construed as ambiguous or lead to unintended outcomes.  Provide ways in which the language can be more precise.''
\end{displayquote}

By providing the prompt above along with the captured requirements, the LLM can provide results of interest to the product manager. For example, if some requirement specifications were:

\begin{enumerate}
    \item Create a user interface that allows file uploads via web browser.
    \item Keep the files secure.
    \item Make sure the files are movie file types.
    \item Files can be no larger than 100 MB
\end{enumerate}

the LLM would provide a path for the product manager to bolster the requirement specifications prior to providing them to the development team. For example the LLM might suggest 
\begin{displayquote}
        ``The system requirement to ``Keep the files secure'' could be made more precise by specifying which security measures should be implemented to ensure the files are kept secure.
        For example: ``Implement access control and encryption measures to ensure that uploaded files can only be accessed by authorized users and are protected against unauthorized access or tampering.''
\end{displayquote}

Follow-up questions such as ``provide example requirements for a system that accepts file uploads by logged-in users, and keeps the files secure the entire time'' may then be posed to guide product manager decision-making and ensure the  requirements delivered to developers are more precise and actionable.

\subsubsection{Consequences}

A difficulty of creating a specification, which could take the form of requirements or non-formal description of the system, is that errors, areas of ambiguity, or omissions are often not discovered until after a system is initially implemented and deployed. In many cases, developers  make assumptions that enable them to implement ambiguous ideas, even if those assumptions do not align with the original specification designers intention. Even experienced system designers and software engineers are prone to ambiguous language and incomplete specification, particularly for ultra large-scale systems~\cite{FeilerUltraLargeScaleSystems2006}. 

The \textit{Specification Disambiguation} pattern helps overcome this issue by providing an automated "devil's advocate" that can attempt to find points of weakness in a requirement specification. In addition, whereas social concerns (such as the concern of questioning a more senior developer) may cause developers to not ask questions, LLMs do not suffer from these same biases and reservations.

A particularly compelling use case for this pattern is in the integration of two separate systems, such as two different development teams building separate parts of a system using a common integration specification. This pattern can be used independently by the team members to collect potential ambiguities and then bring them to a joint meeting before kicking off development to discuss. The LLM can serve as an unbiased source of topics of discussion for a kick-off. 

The \textit{Specification Disambiguation} pattern  pattern is also helpful when a specification is being developed for an external audience (such as consumers of an unreleased API) who are not involved in the specification writing process. In this case, developers may not be able to talk to the target consumers due to secrecy or lack of an audience for the product, and thus lack the mean to easily get external feedback on the specification. This patterns allows the LLM to serve as a representative for external users. 

The pattern is also effective when combined with the \textit{Persona}~\cite{white2023prompt}, \textit{API Generator}, and \textit{API Simulator} or \textit{Requirements Simulator} patterns. Ambiguities can be further discovered by interactively simulating the system or converting it into an API specification. Each transforming the specification into another format through one of these prompt patterns can help identify ambiguities that are uncovered through this pattern since the transformation may produce an unexpected reification of the requirements. In addition, the \textit{Persona} pattern can be used to consider potential ambiguities from different perspectives.

\subsection{The API Generator Pattern}
\label{apigen}

\subsubsection{Intent and Context}
This pattern generates an application programming interface (API) specification, such as a REST API specification, from 
natural language requirement statements or descriptions of the system. The goal of this pattern is to allow developers to rapidly explore multiple possible API designs, formalize API design earlier, and produce a starting point for manual refinement of the design. 

\subsubsection{Motivation} 
Designing a complete API specification to support a set of requirements---or even a high-level description of a software-reliant system---often involves manual effort. If this level of effort is significant then (1) fewer potential designs may be explored, (2) systematic API specifications may be deferred until after key portions of the system are implemented, and/or (3) \textit{ad hoc} alignment and integration of disparate systems, services, or modules many use the code as the only source of truth. A key motivation for applying the \textit{API Generator} pattern is to dramatically reduce and/or eliminate the cost of the API creation so these specifications are created earlier and careful thought goes into their design.

\subsubsection{Structure and Key Ideas}

The fundamental contextual statements are as follows:

\rowcolors{2}{tablerow1}{tablerow2}
\begin{tcolorbox}[enhanced, notitle, clip upper,
       tabularx*={\renewcommand*{\arraystretch}{1.4}}%
                 {>{\raggedright\arraybackslash\hsize=0.1\hsize}X%
              >{\raggedright\arraybackslash\hsize=1.9\hsize}X}]
    \arrayrulecolor{white}
    \textbf{~} & \textbf{API Generator Pattern} \\ \hline
      1. & Using system description X \\ \hline
      2. & Generate an API specification for the system \\ \hline
      3. & The API specification should be in format Y
\end{tcolorbox}

\subsubsection{Example Implementation}

A sample implementation of this pattern showing a prompt to generate an OpenAPI specification, which is a specification for a REST API, is shown below:

\begin{displayquote}
    "Generate an OpenAPI specification for a web application that would implement the listed requirements."
\end{displayquote}

The implementation uses a concrete format for the specification, OpenAPI, and assumes that the requirements for the system were previously discussed. Typically, this prompt pattern is used after a discussion of the requirements or even a simple textual description of a system, such as "a web application for a customer relationship management system". The more detailed the list of requirements, the more accurate the generated API will be, although developers can perform thought experiments and simulations with prompts as simple as "generate an OpenAPI specification for a web application related to cooking." 

\subsubsection{Consequences}
Providing developers with tools to experiment with different API designs from a system description or requirements list is a powerful development tool. The \textit{API Generator} pattern enables developers and/or teams to rapidly create multiple potential API designs and compare/contrast them before selecting their final design.  In contrast, writing these APIs manually is tedious, so developers often only have time to write and explore a limited number of API design options.

Another benefit of this pattern is that developers may choose to write the API specification after the code is implemented because they do not want to spend time specifying the same information twice (\textit{i.e.}, once in the API specification and again in the actual code). By automating API production, developers are incentivized to create API specifications earlier in the design process. Although existing (i.e., non-LLM) tools can generate an API specification from code, they still require the initial production of code. Moreover, tools that can generate skeletons of code from the API specification can be combined with this pattern to accelerate the API implementation compare with writing it manually.

The \textit{API Generator} pattern can be combined effectively with the \textit{API Simulator} pattern described in 
Section~\ref{apisim} to both generate and evaluate the proposed specification. Simulating the API can allow developers to get a sense of the "ergonomics" of the API and evaluate how hard it is to accomplish various tasks in code. The API can also be refactored through the LLM using the \textit{Data-guided Refactoring} pattern described in
Section~\ref{dataguided}.


\subsection{The API Simulator Pattern}
\label{apisim}

\subsubsection{Intent and Context}
This pattern causes the LLM to simulate the API from a specification, thereby enabling developers to interact immediately with an API and test it through a simulation run by the LLM. LLMs possess powerful---and often largely unrecognized---capabilities to generate synthetic data and tailor that data to natural language descriptions of scenarios. In addition, LLMs can help developers explore a simulated API by synthesizing sample requests, as well as providing usage guidance and explaining errors. 

\subsubsection{Motivation} 
Although tools are available to simulate an API~\cite{kendarorgHttpAnsweringMachine}, they require setup to use and may have limited ability to generate effective synthetic data. Current infrastructure for simulating APIs also often just supports strict interaction, typically through HTTP or code, rather than a more fluid interface based on a combination of pseudo-operations and concrete operation specification. Early interaction with an API design can aid developers in uncovering issues, omissions, and awkward designs. 

\subsubsection{Structure and Key Ideas}

The fundamental contextual statements are as follows:

\rowcolors{2}{tablerow1}{tablerow2}
\begin{tcolorbox}[enhanced, notitle, clip upper,
       tabularx*={\renewcommand*{\arraystretch}{1.4}}%
                 {>{\raggedright\arraybackslash\hsize=0.1\hsize}X%
              >{\raggedright\arraybackslash\hsize=1.9\hsize}X}]
    \arrayrulecolor{white}
    \textbf{~} & \textbf{API Simulator Pattern} \\ \hline
      1. & Act as the described system using specification X \\ \hline
      2. & I will type in requests to the API in format Y \\ \hline
      3. & You will respond with the appropriate response in format Z based on specification X
\end{tcolorbox}

\subsubsection{Example Implementation}

An example implementation of this pattern that asks the LLM to simulate a REST API based on an OpenAPI specificaton is shown below. The implementation specifies that requests to the system will be typed in as HTTP requests and that the system should output the HTTP response. It is also possible to have the LLM generate a description of state changes in the system as the simulation, data saved, etc. Similarly, the specification of the user input could be simply a description of what a user is doing with the API or a web client.

\begin{displayquote}
    "Act as this web application based on the OpenAPI specification. I will type in HTTP requests in plain text and you will respond with the appropriate HTTP response based on the OpenAPI specification."
\end{displayquote}

The specification can take a number of forms, such as a programmatic interface or a common API specification domain-specific language, such as
OpenAPI~\cite{openapis}. In the example above, the OpenAPI specification for an HTTP API is used. Requests can then be input to the LLM, which then replies with the corresponding HTTP responses. 

\subsubsection{Consequences}

One benefit of the \textit{API Simulator} pattern is that users can customize their interactions or scenarios of interaction using natural language, which may be easier than trying to accomplish the same thing in code. For example,  users can tell the LLM, "for the following interactions, assume that the user has a valid OAuth authorization token for the user Jill" or "assume that 100 users exist in the system and respond to requests with sample data for them." More complex customization can also be performed, such as "assume the users are from a set of 15-25 families and come from multiple countries in North America." 

Interactions with a simulated API can be done through either a rigorous programmatic form, such as "strictly interpret my input as an HTTP request and reject any malformed requests" or "I am going to only provide pseudo data for the input and you should fix any formatting issues for me." The flexibility of interacting with the LLM simulation and customizing it via natural language facilitates rapid exploration of an API. 

Another key benefit arises when combining the \textit{API Simulator} is that the user can have the LLM create examples of usage that are later used as few-shot examples~\cite{wang2020generalizing} in future prompts or that users can leverage to reason about how hard or easy it is to accomplish various tasks in code. A pattern to combine with is \textit{Change Request Simulation} described in Section~\ref{changerequest}, which allows users to reason about the effort needed to accommodate changing assumptions later in the software life-cycle. 


\subsection{Pattern: Few-shot Code Example Generation}

\subsubsection{Intent and Context} 
The goal of the pattern is to get the LLM to generate a set of usage examples that can later be provided back to the LLM as examples in a prompt to leverage few-shot learning~\cite{wang2020generalizing}. Few-shot learning is based on providing a limited set of example training data in a prompt to an LLM. In the case of code, the few-shot examples are proper usage of the code that the LLM can learn from. In some cases, code examples can convey the function and use of the code in a more space / token-efficient manner than the actual code itself. This pattern leverages the LLM itself to generate few-shot examples that can later be provided in a prompt, in place of the actual code, to allow the LLM to reason about the original code. These examples can be helpful to remind the LLM of the design / usage of the system that it designed in prior conversations. 

\subsubsection{Motivation}
 Since a large software system or module may exceed the token limit of an LLM, a method is needed to describe design aspects to the LLM, such as a module, class, set of functions, etc. within the token limits and how to properly use the system. This overrunning of the token limit necessitates a way to remind the LLM of prior design decisions that it made in the past. One approach to solve this problem is to provide few-shot training examples in a prompt that are based on the usage of the code, API, state transitions, or other specification usage examples. These examples can demonstrate proper usage and train the LLM on the fly to properly use the related design or code. However, manually generating few-shot examples may not be feasible. 

\subsubsection{Structure and Key Ideas}

\rowcolors{2}{tablerow1}{tablerow2}
\begin{tcolorbox}[enhanced, notitle, clip upper,
       tabularx*={\renewcommand*{\arraystretch}{1.4}}%
                 {>{\raggedright\arraybackslash\hsize=0.1\hsize}X%
              >{\raggedright\arraybackslash\hsize=1.9\hsize}X}]
    \arrayrulecolor{white}
    \textbf{~} & \textbf{Few-shot Code Example Generation Pattern} \\ \hline
      1. & I am going to provide you system X \\ \hline
      2. & Create a set of N examples that demonstrate usage of system X \\ \hline 
      3. & Make the examples as complete as possible in their coverage \\ \hline
      4. & (Optionally) The examples should be based on the public interfaces of system X \\ \hline 
      5. & (Optionally) The examples should focus
      on X
\end{tcolorbox}

\subsubsection{Example Implementation}

The example implementation below generates few-shot examples of usage of a REST API and focuses the examples on the creation of new users. These examples could then be used in later prompts to the LLM to remind it of the design of the API with regard to the creation of users. Providing the examples may be more concise and convey more meaning than natural language statements that try to convey the same information.

\begin{displayquote}
    "I am going to provide you code. Create a set of 10 examples that demonstrate usage of this OpenAPI specification related to registration of new users."
\end{displayquote}

In this next example, the pattern implementation asks the LLM to create few-shot examples for usage of portion of code. Code examples can be very information-rich and token-efficient relative to natural language, particularly when the examples convey important meaning, such as ordering of operations, required data, and other details that are concise when described in code but verbose in natural language.

\begin{displayquote}
    "I am going to provide you code. Create a set of 10 examples that demonstrate usage of this code. Make the examples as complete as possible in their coverage. The examples should be based on the public interfaces of the code."
\end{displayquote}

\subsubsection{Consequences}

A key benefit of this pattern is that it can be used early in the design cycle to help capture expected usage of a system and then later provide a usage-based explanation back to the LLM to illustrate its past design decisions. When combined with patterns, such as the \textit{API Simulator} pattern, developers can rapidly interact with the system and record the interactions and then supplement them with additional generated examples. 

This pattern is best applied when example usage of the system also conveys important information about constraints, assumptions, or expectations that would require more tokens to express in another format, such as a written natural langauge description. In some cases, a document, such as an OpenAPI specification, may be more token-efficient for conveying information. However, example usage has been shown to be an effective way of helping an LLM perform problem solving~\cite{yao2022react}, so this pattern may be a useful tool even when it is not the most token-efficient mechanism for conveying the information. 


\subsection{The Domain-Specific Language (DSL) Creation Pattern}

\subsubsection{Intent and Context}
This pattern enables an LLM to create its own domain-specific language (DSL) that both it and users can leverage to describe and manipulate system concepts, such as requirements, deployment aspects, security rules, or architecture in terms of modules. The LLM can then design and describe the DSL to users. In addition, the examples and descriptions the LLM generates can be stored and used in future prompts to reintroduce the DSL to the LLM. Moreover, the examples the LLM generates will serve as 
few-shot examples for future prompts. 

\subsubsection{Motivation} 
DSLs can often be used to describe aspects of a software-reliant system using more succinct and token-efficient formats than natural language, programming languages, or other formats~\cite{md-engineering}. LLMs have a maximum number of "tokens", which corresponds to the maximum size of a prompt, and creating more token-efficient inputs is important for large software projects where all the needed context may be hard to fit into a prompt. Creating a DSL, however, can be time-consuming. In particular, the syntax and semantics of the DSL (\textit{e.g.}, its metamodel) must be described to an LLM \textit{a priori} to enable subsequent interactions with users. 

\subsubsection{Structure and Key Ideas}

The fundamental contextual statements are as follows:

\rowcolors{2}{tablerow1}{tablerow2}
\begin{tcolorbox}[enhanced, notitle, clip upper,
       tabularx*={\renewcommand*{\arraystretch}{1.4}}%
                 {>{\raggedright\arraybackslash\hsize=0.1\hsize}X%
              >{\raggedright\arraybackslash\hsize=1.9\hsize}X}]
    \arrayrulecolor{white}
    \textbf{~} & \textbf{DSL Creation Pattern} \\ \hline
      1. & I want you to create a domain-specific language for X \\ \hline
      2. & The syntax of the language must adhere to the following constraints \\ \hline
      3. & Explain the language to me and provide some examples
\end{tcolorbox}

\subsubsection{Example Implementation}

A sample implementation of this pattern creating a DSL for requirements is shown below. This implementation adds a constraint that the DSL syntax should be YAML-like, which aids the LLM in determining what the textual format should take. An interesting aspect of this is that "like" may yield a syntax that is not valid YAML, but looks similar to YAML.

\begin{displayquote}
    "I want you to create a domain-specific language to document requirements. The syntax of the language should be based on YAML. Explain the language to me and provide some examples."
\end{displayquote}

Another implementation approach is to ask the LLM to create a set of related DSLs with references between them. This approach is helpful when you need to describe related aspects of the same system and want to trace concepts across DSL instances, such as tracing a requirement to its realization in an architectural DSL describing modules. The LLM can be instructed to link the same concept together in the DSLs through consistent identifiers so that concepts can be tracked across DSL instances.

\subsubsection{Consequences}

One consequence of the \textit{DSL Creation} pattern is that it may facilitate system design without hitting 
token limits. The specific syntax rules that are given to the LLM must be considered carefully, however, since they directly influence the space-efficiency of the generated DSL. Although users of a DSL may only need to express relevant concepts for a designated task, this high 
concept-density may not translate into the token-efficiency of a textual representation of such concepts. For example, an XML-based syntax for a DSL will likely be much more space consumptive than a YAML-based syntax. 

Token efficiency in a DSL design can be improved via conventions and implicit syntax rules. For example, positional conventions in a list can add meaning rather than marking different semantic elements in the list via explicit labels. The downside, however, is that the DSL may be harder to interpret for users unfamiliar with its syntax, although this problem can be rectified by using the \textit{Few-shot Code Example Generation} 
pattern to create examples that teach users how to apply the DSL. Combining these two pattern also helps the LLM self-document usage of the pattern for later prompting based on the DSL.


\subsection{The Architectural Possibilities Pattern}

\subsubsection{Intent and Context}
This pattern generates several different architectures for developers to consider, with little effort on the part of developers. An "architecture" can be very open-ended and it is up to the developer to explain to the LLM what is meant by this term. A developer may desire seeing alternative architectures for how code is laid out into files, communication is performed between modules, or tiers in a multi-tiered system. The intent of the pattern is to allow the developer to explore any of these architectural aspects of the system with the LLM. Moreover, developers can interactively refine architectural suggestions by adding further constraints or asking the LLM to describe the architecture in terms of a different aspect of the system, such as file layout, modules, services, communication patterns, infrastructure, etc.

\subsubsection{Motivation}
Devising software architectures often requires considerable cognitive effort on the development team, particularly when  architectures are mapped all the way to system requirements. Developers may therefore only consider a relatively small number of possible architectures when designing a software-reliant system due to the effort required to generate such architecture. In addition, developers may not have familiarity with architectures that could be a good fit for their systems and hence would not explore these architectural possibilities. Since architecture plays such an important role in software-reliant system design, it is important to facilitate exploration of many different alternatives, including alternatives that developers many not be familiar with. 

\subsubsection{Structure and Key Ideas}
The fundamental contextual statements are as follows:

\rowcolors{2}{tablerow1}{tablerow2}
\begin{tcolorbox}[enhanced, notitle, clip upper,
       tabularx*={\renewcommand*{\arraystretch}{1.4}}%
                 {>{\raggedright\arraybackslash\hsize=0.1\hsize}X%
              >{\raggedright\arraybackslash\hsize=1.9\hsize}X}]
    \arrayrulecolor{white}
    \textbf{~} & \textbf{Architectural Possibilities Pattern} \\ \hline
      1. & I am developing a software system with X for Y \\ \hline
      2. & The system must adhere to these constraints \\ \hline
      3. & Describe N possible architectures for this system \\ \hline 
      4. & Describe the architecture in terms of Q
\end{tcolorbox}

\subsubsection{Example Implementation}

The example implementation below explores architectures related to using a web application built on a specific set of frameworks:

\begin{displayquote}
    "I am developing a python web application using FastApi that allows users to publish interesting ChatGPT prompts, similar to twitter. Describe three possible architectures for this system. Describe the architecture with respect to modules and the functionality that each module contains."
\end{displayquote}

The implementation specifies that the architecture should be described in terms of the modules and functionality within each module. The "with respect to" portion of the pattern is important to guide the LLM's output to appropriately interpret the term architecture. The same prompt could be changed to ask for architecture in terms of the REST API, interaction of a set of services, communication between modules, data storage, deployment on virtual machines, or other system aspects. The "with respect to" focuses the output on which of the many aspects the architecture is being explored in terms of.

\subsubsection{Consequences}

Performance-sensitive applications can use this pattern to propose possible architectures to meet performance goals and then generate experiments, in the form of code, to test each architecture. For example, a cloud application might be implementable as 
(1) a monolithic web application and run in a container or (2) a series of services in a microservice architecture. The LLM can first generate a sample implementation of each architecture and then generate a script to deploy each variation to the cloud and test it under various workloads. In addition, the workload tests could allow for comparative cost analysis from the resulting expenses incurred in the cloud. The \textit{Architectural Possibilities} pattern is particularly powerful when combined with this type of LLM-based rapid implementation and experimentation. 

Another way to expand this rapid architectural experimentation capability is to combine it with the \textit{API Generator} pattern described in 
Section~\ref{apigen} and \textit{API Simulator} pattern described in Section~\ref{apisim}. The architecture can serve as the basis of the API generation, which can then be simulated. This approach allows developers to see what the realization and use of this architecture from a code-perspective might look like. Likewise, the \textit{Change Request Simulator} pattern described in Section~\ref{changerequest} can be employed to reason about how hard/easy it would be to change different assumptions later given a proposed architecture.


\subsection{The Change Request Simulation Pattern}
\label{changerequest}

\subsubsection{Intent and Context}
This pattern helps users reason about the complexity of a proposed system change, which could be related to requirements, architecture, performance, etc. For example, this pattern helps  users reason about what impact a given change might have on some aspect of the system, such as which modules might need changing. This pattern is particularly helpful when a group of stakeholders need to discuss a possible requirements change and the LLM can serve as a (potentially) unbiased estimator of the scope and impact of the change.

\subsubsection{Motivation}
In many situations, it may not be immediately clear to stakeholders what the impact of a change would be. Without an understanding of the impact of a change, it is hard to reason about the associated effects on schedule, cost, or other risks. Getting rapid feedback on potential impacts can help stakeholders initiate the appropriate conversations and experiments to better determine the true risk of the change. Distrust between users may also complicate the discussion of the change and necessitate an ``unbiased'' external opinion.

\subsubsection{Structure and Key Ideas}
The fundamental contextual statements are as follows:

\rowcolors{2}{tablerow1}{tablerow2}
\begin{tcolorbox}[enhanced, notitle, clip upper,
       tabularx*={\renewcommand*{\arraystretch}{1.4}}%
                 {>{\raggedright\arraybackslash\hsize=0.1\hsize}X%
              >{\raggedright\arraybackslash\hsize=1.9\hsize}X}]
    \arrayrulecolor{white}
    \textbf{~} & \textbf{Change Request Simulation Pattern} \\ \hline
      1. & My software system architecture is X \\ \hline
      2. & The system must adhere to these constraints \\ \hline
      3. & I want you to simulate a change to the system that I will describe \\ \hline 
      4. & Describe the impact of that change in terms of Q \\ \hline
      5. & This is the change to my system
\end{tcolorbox}

\subsubsection{Example Implementation}

In this example implementation, the prompt refers back to a previously generated OpenAPI specification as the basis of the simulation:

\begin{displayquote}
    "My software system uses the OpenAPI specification that you generated earlier. I want you to simulate a change where a new mandatory field needs to be added to the prompts. List which functions and which files will need to be modified."
\end{displayquote}

The implementation focuses the simulation on how the change will impact various functions and files in the system. This approach would allow the developer to estimate the cost of a change by looking at the complexity of the referenced files, functions, and the total count of each. Alternatively, in cases where the entire affected section of code can fit into the prompt, the LLM can be asked to identify lines of code that may need changing. 

\subsubsection{Consequences}

The hardest part of applying the \textit{Change Request Simulation} pattern is establishing enough context for the LLM to reason about a proposed change. This pattern works best when it is employed with other System Design category patterns, such as the \textit{API Generator}, where the conversation history can be used to seed the analysis. The more concrete the change description is in relation to the context, the more likely the LLM can provide a reasonable estimate of change impact.

This pattern can also be used to reason either (1) abstractly about a software-reliant system in terms of modules or (2) concretely in terms of files, functions, and/or lines of code. Existing LLMs have token limits that only consider a limited amount of information about a system. Large sweeping changes to a system can generally only be reasoned about at a higher level of abstraction since the detailed information needed to implement such changes would exceed the token limit of the LLM. Within a smaller set of files or feature, however, an LLM may be able to reason precisely about what needs to change. 

One way to handle the tension between token limits and detailed  output is to apply the \textit{Change Request Simulation} pattern iteratively to zoom in and out. Initially, an abstract analysis is performed to identify features, modules, etc. that need to change. The prompt is then modified to refine the context to a specific module or feature and obtain greater detail from the LLM. The process can be repeated on individual parts of the module or feature until sufficient detail is gained. This process can be repeated for each high-level module identified to gather an estimate of the overall impact of a proposed change. 


\section{Code Quality \& Refactoring Patterns}
\label{qual}

This section describes prompt patterns that help LLMs generate higher quality code and refactor human-produced code.

\subsection{The Code Clustering Pattern}

\subsubsection{Intent and Context}
The goal of this pattern is to separate and cluster code into functions, classes, etc. based on a particular property of the code, such as separating pure and impure code, business logic from database access, HTTP request handling from business logic. etc. The pattern defines the expected cluster properties to the LLM and then asks the LLM to automatically restructure the code to realize the desired clustering. The pattern can be used to ensure that LLM-generated code exhibits the clustering, refactor human-produced code to add the clustering, or review code.

\subsubsection{Motivation} 
How code is sliced up and clustered into functions, modules, etc. has a significant impact on what can be changed, extended, and easily maintained. By default, an LLM will not have guidelines on the clustering / slicing needs for the application. This lack of context can lead an LLM to produce code that is perceived as monolithic, brittle, messy, and overall poor in quality. The motivation is to provide the missing clustering context that the LLM needs to produce code that is higher quality.

\subsubsection{Structure and Key Ideas}

The fundamental contextual statements are as follows:

\rowcolors{2}{tablerow1}{tablerow2}
\begin{tcolorbox}[enhanced, notitle, clip upper,
       tabularx*={\renewcommand*{\arraystretch}{1.4}}%
                 {>{\raggedright\arraybackslash\hsize=0.1\hsize}X%
              >{\raggedright\arraybackslash\hsize=1.9\hsize}X}]
    \arrayrulecolor{white}
    \textbf{~} & \textbf{Code Clustering Pattern} \\ \hline
      1. & Within scope X \\ \hline
      2. & I want you to write or refactor code in a way that separates code with property Y from code that has property Z. \\ \hline
      3. & These are examples of code with property Y. \\ \hline
      4. & These are examples of code with property Z. 
\end{tcolorbox}

\subsubsection{Example Implementation}

A sample implementation of this pattern is shown below:

\begin{displayquote}
    "Whenever I ask you to write code, I want you to write code in a way that separates functions with side-effects, such as file system, database, or network access, from the functions without side-effects."
\end{displayquote}

One way to specify the properties that are being used for subdivision is by defining one property as the absence of the other property. In the example above, the "side-effects" property is clearly defined. The "without side-effects" property is defined as the converse of the side-effects property. A common form of implementation is to define properties that are opposites of each other. 

Some common properties that are effective in generating higher quality code with this pattern include: 
\begin{itemize}
    \item Side-effects: isolate side-effects so that the code is easier to test, business logic can be reasoned about in isolation, etc.
    \item Tiers: isolate code based on a layered architecture, such as business and data tiers
    \item Features: group code into cohesive features that are isolated in separate files or groups of files
\end{itemize}

Many other properties can be used to separate and cluster code as long as they are describable to the LLM. Well-understood properties, such as side-effects, are likely to have been concepts that were present in the LLM's training set. These types of properties will require less prompt design work for the LLM to reason about. Custom properties can be reasoned about through a combination of natural language description and few-shot examples. This pattern can be combined with the \textit{Few-shot Code Example Generator} pattern to create code samples that demonstrate the desired property-based clustering and then use them in this pattern for in-context learning of the property.

\subsubsection{Consequences}

This is one of the most basic patterns that can dramatically improve the perceived quality of LLM-produced code. Unless the LLM is told otherwise, its code will solve the problem at hand and often does not solve structuring problems, such as separating pure and impure functions, that it has not been asked to solve. The pattern surfaces a key issue in LLM software engineering, which is, an LLM's output is only as good as the prompt it is given. Implicit knowledge, such as that the project requires code that illustrates certain clustering properties, will not be known to the LLM unless it is provided this information in a prompt. 

\subsection{The Intermediate Abstraction Pattern}

\subsubsection{Intent and Context}

Abstraction and modularity are fundamental components of high-quality maintainable and reusable code. Code should be written in a way that isolates cohesive concepts into individual functions or classes so that edits can be isolated in scope. In addition, when working with an LLM, refactoring existing code is easier if the refactorings can be isolated to a single function that needs to be modified, replaced, or added. 

\subsubsection{Motivation} 

Be default, LLMs will generate code that is very procedural and directly translates the goals into code. However, the implementation may or may not have any reasonable abstraction or modularity, leading it to be difficult to maintain. Further, as the LLM is asked to continually add features to the code, it may produce longer and longer functions with little to no separation of concepts into modules, functions, or other abstractions that will facilitate long-term maintainability.

\subsubsection{Structure and Key Ideas}

The fundamental contextual statements are as follows:

\rowcolors{2}{tablerow1}{tablerow2}
\begin{tcolorbox}[enhanced, notitle, clip upper,
       tabularx*={\renewcommand*{\arraystretch}{1.4}}%
                 {>{\raggedright\arraybackslash\hsize=0.1\hsize}X%
              >{\raggedright\arraybackslash\hsize=1.9\hsize}X}]
    \arrayrulecolor{white}
    \textbf{~} & \textbf{Intermediate Abstraction Pattern} \\ \hline
      1. & If you write or refactor code with property X \\ \hline
      2. & that uses other code with property Y \\ \hline
      3. & (Optionally) Define property X \\ \hline
      4. & (Optionally) Define property Y \\ \hline
      5. & Insert an intermediate abstraction Z between X and Y \\ \hline
      6. & (Optionally) Abstraction Z should have these properties
\end{tcolorbox}

\subsubsection{Example Implementation}

A sample implementation of this pattern is shown below:

\begin{displayquote}
    "Whenever I ask you to write code, I want you to separate the business logic as much as possible from any underlying 3rd-party libraries.  Whenever business logic uses a 3rd-party library, please write an intermediate abstraction that the business logic uses instead so that the 3rd-party library could be replaced with an alternate library if needed."
\end{displayquote}

A common risk in software is 3rd-party libraries. Since 3rd-party library dependencies are not directly under the control of a developer, they can create risk in a project. For example, the developer of the dependency may make breaking changes to the dependency that make it difficult to incorporate into an existing project and limit access to future security updates in the new version. This example implementation uses the LLM to explicitly insert an intermediate abstraction to protect against this type of risk. 

\subsubsection{Consequences}

One consideration of this pattern is that it may not be possible to design a good abstraction simply from analysis of a single 3rd-party library that provides a given capability. For example, different dependencies may have different fundamental architectures and interfaces. One way around this is to leverage the \textit{Few-shot Example Generator} pattern to create examples of other comparable 3rd-party libraries and their usage and then ask the LLM to refactor the interface to be implementable with any of the alternatives.


\subsection{The Principled Code Pattern}

\subsubsection{Intent and Context}
 The goal of the pattern is to use well-known names for coding principles to describe the desired code structure without having to explicitly describe each individual design rule. For example, an organization may want to ensure that their code follows SOLID design principles. The goal is to quickly ensure that generated, refactored, and reviewed code adheres to expected design or other principles. 

\subsubsection{Motivation}
Writing code with good design characteristics is important to providing maintainable code, but a developer may not be able to easily specify all of the specific rules for what constitutes good design. Long articles and papers are written that explain how to apply good design practices to different languages, frameworks, etc. The motivation for this pattern is that it may be beyond a developer's capabilities to define these rules in natural language, even if they know the fundamental name of the design methodology.

\subsubsection{Structure and Key Ideas}

The fundamental contextual statements are as follows:

\rowcolors{2}{tablerow1}{tablerow2}
\begin{tcolorbox}[enhanced, notitle, clip upper,
       tabularx*={\renewcommand*{\arraystretch}{1.4}}%
                 {>{\raggedright\arraybackslash\hsize=0.1\hsize}X%
              >{\raggedright\arraybackslash\hsize=1.9\hsize}X}]
    \arrayrulecolor{white}
    \textbf{~} & \textbf{Principled Code Pattern} \\ \hline
      1. & Within this scope \\ \hline
      2. & Generate, refactor, or create code to adhere to named Principle X
\end{tcolorbox}

\subsubsection{Example Implementation}

A sample implementation of the prompt pattern is shown below:

\begin{displayquote}
    "From now on, whenever you write, refactor, or review code, make sure it adheres to
    SOLID design principles."
\end{displayquote}

This example uses the SOLID design principles~\cite{marshall2009solid} as the desired design guidelines. SOLID code refers to code that follows five explicit design principles: 1) single responsibility, 2) open-closed, 3) Liskov substitution, 4) interface segregation, and 5) dependency inversion. The named design methodology immediately introduces the underlying five principles that the code should follow.  

\subsubsection{Consequences}

This pattern works best when there is a substantial volume of written material that the LLM was trained on that explains the application of the named principle to different code bases. The more well-known the design principle, the more examples the LLM will likely have been trained on. The availability of training examples is particularly important for less mainstream languages or languages with more uncommon designs, such as Prolog.  

This is similar to the \textit{Persona Pattern}~\cite{white2023prompt} outlined in our prior work, where the user describes the desired output using a well-known name. A key consequence of this pattern is that it will only work with well-known named descriptions of code qualities from prior to when the LLM was trained. Newer coding or design styles that came after the training date will not be accessible through this pattern. However, other approaches could be used to leverage in-context learning and few-shot examples to illustrate these inaccessible named coding or design styles.


\subsection{The Hidden Assumptions Pattern}

\subsubsection{Intent and Context}
The goal of this pattern is to have the LLM identify and describe any assumptions that are made in a section of code. The pattern helps the user identify these assumptions or remind them of assumptions that they may have forgotten about. By showing key assumptions from the code to users, they LLM can help make sure that the user accounts for these assumptions in their decisions related to the code.

\subsubsection{Motivation}
Any code, regardless if it is produced by a human or LLM, may have hidden assumptions that the user needs to understand. If the user is not aware of these assumptions, they may use, modify, or otherwise leverage the code incorrectly. This is a particular risk for LLM-generated code, where the user has less familiarity with what is being produced for them. 

\subsubsection{Structure and Key Ideas}

The fundamental contextual statements are as follows:

\rowcolors{2}{tablerow1}{tablerow2}
\begin{tcolorbox}[enhanced, notitle, clip upper,
       tabularx*={\renewcommand*{\arraystretch}{1.4}}%
                 {>{\raggedright\arraybackslash\hsize=0.1\hsize}X%
              >{\raggedright\arraybackslash\hsize=1.9\hsize}X}]
    \arrayrulecolor{white}
    \textbf{~} & \textbf{Hidden Assumptions Pattern} \\ \hline
      1. & Within this scope \\ \hline
      2. & List the assumptions that this code makes \\ \hline
      3. & (Optionally) Estimate how hard it would be to change these assumptions or their likelyhood of changing
\end{tcolorbox}

\subsubsection{Example Implementation}

A sample implementation of this pattern is shown below:

\begin{displayquote}
    "List the assumptions that this code makes and how hard it would be to change each of them given the current code structure."
\end{displayquote}

This first example focuses on listing assumptions that may be hard to change in the future. This is a refinement of the pattern that helps make the developer aware of what the liabilities in the code may be with respect to future change. If one of the assumptions isn't easy to change, but the developer expects that this aspect will need to change, they can request a refactoring by the LLM to remove this flawed assumption.

A second example of this pattern shows how it can be used to aid in refactoring code away from being tightly-coupled to an underlying database.

\begin{displayquote}
    "List the assumptions in this code that make it difficult to change from a MongoDB database to MySQL."
\end{displayquote}

With this example, the LLM will list the assumptions that are tightly coupling to a specific database. The user could then take this list and use it as the basis of a refactoring, by asking the LLM to refactor the code to eliminate the listed assumptions.

\subsubsection{Consequences}

One consequence of this pattern is that it may not identify all of the assumptions in the code. For example, there may be code outside of what is in the context provided to the LLM that would be needed to identify the assumption. The risk of this pattern is that developers will take it as a source of truth for all assumptions in the code rather than flagging of some possible assumptions for the developer's consideration. 


\subsection{The Pseudo-code Refactoring Pattern}

\subsubsection{Intent and Context}
The goal of this pattern is to give the user more fine-grained control over the algorithm, flow, or other aspects of the code, while not requiring explicit specification of details. The pattern allows the user to define pseudo-code for one or more details of the generated or refactored code. The LLM is expected to adapt the output to fit the pseudo-code template while ensuring that the code is correct and runnable. 

\subsubsection{Motivation}
In many cases, a user may have strong opinions or specific goals in the refactoring or generation of code that would be tedious to describe (and duplicative of the LLM's work) if they required typing in the exact code structures they wanted. If the developer has to do as much coding work as the LLM to specify what they want, the benefit of using an LLM is reduced. The motivation of the pattern is to provide a middle ground that allows greater control over code aspects without explicit coding and consideration of details.

\subsubsection{Structure and Key Ideas}

Fundamental contextual statements:

\rowcolors{2}{tablerow1}{tablerow2}
\begin{tcolorbox}[enhanced, notitle, clip upper,
       tabularx*={\renewcommand*{\arraystretch}{1.4}}%
                 {>{\raggedright\arraybackslash\hsize=0.1\hsize}X%
              >{\raggedright\arraybackslash\hsize=1.9\hsize}X}]
    \arrayrulecolor{white}
    \textbf{~} & \textbf{Pseudo-code Refactoring Pattern} \\ \hline
      1. & Refactor the code \\ \hline
      2. & So that it matches this pseudo-code \\ \hline 
      3. & Match the structure of the pseudo-code as closely as possible
\end{tcolorbox}

\subsubsection{Example Implementation}

A sample implementation of this pattern is shown below:

\begin{displayquote}
    "Refactor the following code to match the following psuedo-code. Match the structure of the pseudo-code as closely as possible. 
    \begin{verbatim}
    files = scan_features()
    for file in files:
        print file name
    for file in files:
        load feature
        mount router
    create_openapi()
    main():
       launch app"        
    \end{verbatim}

\end{displayquote}

In this example, the user is asking to have a much larger body of code refactored to match the structure of the pseudo-code. The pseudo-code defines the outline of the code, but not the details of how the individual tasks are accomplished. In addition, the pseudo-code does not provide exact traceability to which lines are expected to be part of the described functionality. The LLM figures out what the intent of the refactoring is and how to map it into the current code provided to it.

\subsubsection{Consequences}

This pattern can lead to more substantial refactoring than what is outlined in the pseudo-code. For example, rewriting the code to match the pseudo-code may require removing a function and splitting its code between two other functions. Removal of the function could then change the public interface of the code. 


\subsection{The Data-guided Refactoring Pattern}
\label{dataguided}

\subsubsection{Intent and Context}
The goal of this pattern is to allow the user to refactor existing code to use data with a new format. Rather than specifying the exact logic changes to be able to use the new format, the user can provide the new format schema to the LLM and ask the LLM to determine how to make the necessary changes. The pattern helps to automate code refactoring for this common task of incorporating changes to data formats.

\subsubsection{Motivation}
Refactoring code to use a new input or output data structure can be tedious. When communicating with an LLM, explaining the explicit refactoring steps to may also require more time than actually conducting the needed steps. Therefore, a concise way of explaining to the LLM the desired refactoring is needed.

\subsubsection{Structure and Key Ideas}

Fundamental contextual statements:

\rowcolors{2}{tablerow1}{tablerow2}
\begin{tcolorbox}[enhanced, notitle, clip upper,
       tabularx*={\renewcommand*{\arraystretch}{1.4}}%
                 {>{\raggedright\arraybackslash\hsize=0.1\hsize}X%
              >{\raggedright\arraybackslash\hsize=1.9\hsize}X}]
    \arrayrulecolor{white}
    \textbf{~} & \textbf{Data-guided Refactoring Pattern} \\ \hline
      1. & Refactor the code \\ \hline
      2. & So that its input, output, or stored data format is X \\ \hline 
      3. & Provide one or more examples of X
\end{tcolorbox}

\subsubsection{Example Implementation}

An example implementation of the pattern is shown below:

\begin{displayquote}
    "Let's refactor execute\_graph so that graph has the following format \{'graph':\{ ...current graph format... \}, 'sorted\_nodes': \{ 'a': ['b','c'...],...\}\}"
\end{displayquote}

This example asks the LLM to refactor a function to use a different format for the graph. In the example, the specific use of the graph format is not defined, but could potentially be input, output, or internal to the function. All of the different uses of the graph would be supportable by the pattern. In addition, the implementation uses ellipses to indicate portions of the data structure, which allows the LLM to fill in the intent of the user with concrete details. 

\subsubsection{Consequences}
The primary consequence of this pattern is that it reduces the manual effort to specify refactorings for many types of code changes necessitated by a change in data format. In many cases, the refactoring can be completely automated through this process, or at least boostrapped, speeding up and potentially reducing the cost of changing data formats. Since changing data formats can have such a large-scale impact on a system, using this pattern to automate these refactorings can potentially reduce costs and speed up overall system development.

\section{Related Work}
\label{related}

Software patterns~\cite{gamma1995design,schmidt2013pattern} have been studied extensively and shown their value in software engineering. Software design patterns have also been specialized for other types of non-traditional uses, such as designing smart contracts~\cite{DBLP:journals/corr/ZhangWSL17, xu2018pattern}. Prompt design patterns for software engineering are complementary to these, although not focused on the design of the system itself, but on the interactions with an LLM to produce and maintain software-reliant systems over time. 

Prompt engineering is an active area of study and the importance of prompts is well understood~\cite{van2023chatgpt}. Many problems cannot be solved by LLMs unless prompts are structured correctly~\cite{frieder2023mathematical}. Some work has specifically looked at prompting approaches to help LLMs learn to leverage outside tooling~\cite{yao2022react}. Our work complements these approaches, focusing on specific patterns of interaction that can be used to tap into LLM capabilities to solve specific problems in software engineering. 

Much discussion on LLM usage in software engineering to date has centered on the use of LLMs for code generation and the security and code quality risks associated with that usage. For example, Asare et al.~\cite{asare2022github} compared LLM code generation to humans from a security perspective. Other research has examined the quality of generated answers and code from LLMs~\cite{borji2023categorical, frieder2023mathematical,jalil2023chatgpt,nair2023generating} and interaction patterns for fixing bugs~\cite{xia2023conversational,sobania2023analysis}. Our research draws inspiration from these explorations and documents specific patterns that can be used to improve code quality and help reduce errors. Moreover, as more prompt patterns are developed, different patterns can be quantitatively compared to each other for effectiveness in solving code quality issues.

\section{Concluding Remarks}
\label{conclusion}
Much attention has focused on the mistakes that LLMs make when performing software engineering tasks~\cite{github,asare2022github,pearce2022asleep}. As shown in this paper, however, prompt patterns can be used to help combat these mistakes and reduce errors. Moreover, prompt patterns can tap into LLM capabilities, such as simulating a system based on requirements, generating an API specification, pointing out assumptions in code, etc. that are hard to automate using  existing technologies. 

The following are lessons learned thus far from our work on applying ChatGPT to automate common software engineering tasks:

\begin{itemize}

\item \textbf{The depth of capabilities of LLMs, such as ChatGPT, is not widely or fully understood or appreciated}. 
LLMs hold immense potential for helping to automate common tasks throughout the software engineering life-cycle. Many LLM capabilities have the potential to accelerate software engineering, not just by generating code, but by making rapid experimentation at many different levels of abstraction possible. A key to leveraging these capabilities is to codify an effective catalog of prompts and guidance on how to combine them at different stages of the software life-cycle to improve software engineering.

\item \textbf{Significant human involvement and expertise is currently necessary to leverage LLMs effectively for automating common software engineering tasks}.  The tendency of ChatGPT to "hallucinate" confidently and enthusiastically when generating incorrect output requires close scrutiny from human users at this point.  While prompt patterns can help mitigate some of these issues, much further work is needed on other aspects of prompt engineering (such as quality assurance and versioning) to ensure output of LLMs is accurate and helpful in practice.


\end{itemize}

We encourage readers to test the prompt patterns described in this paper by using ChatGPT to replicate our findings in their own domains and environments.

\bibliographystyle{IEEEtran}

\end{document}